\documentstyle[aps,prd]{revtex}
\tighten

\begin{document}

\def \ts {\textstyle}
\def \rd {\displaystyle{\cdot}}
\def \D {\mbox{D}}
\def \ep {\varepsilon}
\def \la {\langle}
\def \ra {\rangle}
\def \c {\mbox{curl}\,}
\def \p {\partial}
\def \cs {c_{\rm s}^2}

\title{Density perturbations with relativistic
thermodynamics}

\author{Roy Maartens\thanks{email: maartens@sms.port.ac.uk}}

\address{School of Computer Science and Mathematics, Portsmouth
University, Portsmouth PO1~2EG, Britain}
        
\author{Josep Triginer\thanks{email: pep@ulises.uab.es}}

\address{Departament de F\'{\i}sica, Universitat Aut\`onoma de
Barcelona, 08193~Bellaterra, Spain}

\maketitle

\begin{abstract}

We investigate cosmological density perturbations in a covariant and 
gauge-invariant formalism, incorporating relativistic 
causal thermodynamics to give a self-consistent description. 
The gradient of density inhomogeneities 
splits covariantly into a scalar part, equivalent to the usual
density perturbations, a rotational vector
part that is determined by the vorticity,
and a tensor part that describes the shape. We give the evolution 
equations for these parts in the general dissipative case. 
Causal thermodynamics gives evolution equations for 
viscous stress and heat
flux, which are coupled to the density perturbation equation
and to the entropy and temperature perturbation equations.
We give the full coupled system in the general dissipative case,
and simplify the system in certain cases.
A companion paper uses the general formalism to analyze 
damping of density perturbations
before last scattering.


\end{abstract}

\pacs{98.80.Hw, 04.40.Nr, 05.70.Ln}

\section{Introduction}

The analysis of density perturbations in cosmological fluids is well 
established, particularly using Bardeen's gauge invariant 
formalism \cite{b,mfb}. This formalism is inherently linear (i.e., it
starts from the background and perturbs away from it) and
non-local. An alternative approach, developed by Ellis and Bruni
\cite{eb}, is covariant (and therefore local) and readily incorporates
nonlinear effects (since it starts from the real spacetime, not the
background). We will use this covariant and gauge-invariant formalism,
in which the variables have a clear physical and geometric
interpretation. Furthermore, the covariant approach is directly
compatible with
causal relativistic thermodynamics, as developed by
Israel and Stewart \cite{is}.

Although dissipative terms representing viscosity and heat conduction
have been formally incorporated into the equations in both
approaches \cite{b,bde}, most applications of the theory are 
restricted to the non-dissipative case -- and even in this case,
relativistic thermodynamics is usually not applied to analyze the
behavior of the fluid self-consistently.
This is is not a problem when studying the evolution of
large-scale perturbations, which are unaffected by local physics --
although the generation of these perturbations,
their initial evolution before leaving 
the Hubble radius, and their final evolution after
re-entering the Hubble radius, are governed by local physics.
For small-scale perturbations, within the Hubble radius, a 
self-consistent analysis requires the application of 
thermodynamics.\footnote{We are considering
here the case of hydrodynamics. Dissipative effects on the microwave
background have been self-consistently analyzed
via numerical integration of
the Boltzmann equation (see, e.g., \cite{hs}). A covariant
and gauge-invariant approach
to the Boltzmann equation is developed in \cite{egs}.}

Here and in a companion paper \cite{tm}, we develop and apply 
a covariant and gauge-invariant analysis of density perturbations
that self-consistently incorporates relativistic 
causal thermodynamics.
The general evolution equations governing density inhomogeneities
are considered in Sec. II. Inhomogeneities are covariantly
characterized by a scalar part, which represents the usual density
perturbations, a vector part, which we show is 
determined by the vorticity, and a tensor part, which determines 
the shape of gravitational clustering.
New evolution equations are derived
for the vector and tensor parts, as well as for perturbations in
the number density, entropy and temperature.
We use the Gibbs equation 
to incorporate the temperature and entropy self-consistently,
and we covariantly characterize different types of perturbation.
In Sec. III, the viscous stress 
and heat flux that appear in the perturbation evolution equations
are subject to thermodynamical 
transport equations, which then form a
coupled system with the perturbation evolution equations.
We define appropriate dissipative scalars to obtain a closed
system of dynamical equations.
The equations are simplified in the particular cases 
of entropy perturbations (non-dissipative), and when only one
form of dissipation is present.

The Israel-Stewart transport equations are 
under reasonable conditions causal and stable
\cite{hl}, and thus provide a consistent relativistic description
of local physical effects on small-scale perturbations.
The thermodynamics of Eckart (and a similar alternative due to
Landau and Lifshitz) is more established in
the literature.
However, in this theory the transport equations reduce 
from evolution equations to algebraic constraints
on viscosity and heat flux, and as a result, the theory
is non-causal (dissipative effects propagate at super-luminal
speeds), and all its equilibrium states are unstable \cite{hl}.
It can be argued \cite{d} that these pathologies only
occur outside the hydrodynamic regime. But firstly, the
stability problem persists in all situations, and secondly, it
seems preferable to employ a theory with built-in causality
and stability. Furthermore, the causal theory can deal with
transient and short-wavelength effects, which are important
in many cosmological and astrophysical situations (see, e.g.,
\cite{caus,his,mm,z}).

Applications to dissipative situations are treated in a
companion paper 
\cite{tm}, where we analyze viscous damping of density perturbations
before last scattering.
This generalizes the results 
of Weinberg \cite{w}, who used non-causal Eckart thermodynamics.

\section{Covariant approach to dissipative perturbations}

The covariant and gauge-invariant analysis of density perturbations
is fully discussed in \cite{eb,bde}. 
Here we present only the main points 
that are necessary for our purposes, before going further by
deriving new evolution equations and 
incorporating causal thermodynamics. Our notation and conventions
follow \cite{bde,m2,mes}, with some changes (see \cite{not}).
 
Given a covariantly defined fluid four-velocity $u^a$ (see the further
discussion below), then $h_{ab}=g_{ab}+u_au_b$ projects into the
local rest spaces of comoving observers, where $g_{ab}$ is the
spacetime metric. The covariant $1+3$ splitting of the Bianchi
identities and the Ricci identity for $u^a$, incorporating
Einstein's field equations as an algebraic definition of the Ricci
tensor, $R_{ab}=T_{ab}-{\ts{1\over2}}Tg_{ab}$, are the fundamental
equations in the covariant perturbation approach. These equations
may be written as propagation and constraint equations for covariant
scalars, spatial vectors ($V_a=h_a{}^bV_b$) and spatial 2-tensors
which are symmetric and trace-free, i.e. which satisfy
\[
S_{ab}=S_{\la ab\ra}\equiv h_a{}^ch_b{}^dS_{(cd)}-{\ts{1\over3}}
h_{cd}S^{cd}h_{ab}\,.
\]
Any spatial 2-tensor has the covariant irreducible decomposition
\[
S_{ab}={\ts{1\over3}}Sh_{ab}+S_{\la ab\ra}+\ep_{abc}S^c\,,
\]
where $S\equiv h^{ab}S_{ab}$ is the spatial trace and
$S_a={\ts{1\over2}}\ep_{abc}S^{bc}$ is the spatial dual to the
skew part. Here $\ep_{abc}=\eta_{abcd}u^d$ is the spatial
permutation tensor defined by projecting the spacetime permutation
tensor $\eta_{abcd}$.
The covariant derivative $\nabla_a$ splits into
a covariant time derivative $\dot{A}_{a\cdots}=
u^b\nabla_bA_{a\cdots}$, and a covariant spatial
derivative $\D_bA_{a\cdots}=h_b{}^dh_a{}^c\cdots
\nabla_dA_{c\cdots}$. (Note that $\D_ch_{ab}=0=\D_d\ep_{abc}$.)
Then the covariant spatial divergence and curl are defined
by \cite{m2}
\begin{eqnarray*}
\mbox{div}\,V=\D^aV_a\,, &~~& \c V_a=\ep_{abc}\D^bV^c\,,  \\
(\mbox{div}\,S)_a=\D^bS_{ab}\,, &~~& 
 \c S_{ab}=\ep_{cd(a}\D^cS_{b)}{}^d\,.
\end{eqnarray*}

The fluid kinematics are described by 
the scalar $\theta=\D^au_a$ (expansion), the spatial vectors
$\dot{u}_a$ (four-acceleration) and
$\omega_a=-{\ts{1\over2}}\c u_a$ (vorticity), and the tensor
$\sigma_{ab}=\D_{\la a}u_{b\ra}$ (shear).
The locally free gravitational
field is described by the electric and magnetic parts of the
Weyl tensor, $E_{ab}=C_{acbd}u^cu^d=E_{\la ab\ra}$ and 
$H_{ab}={\ts{1\over2}}\ep_{acd}C^{cd}{}{}_{be}u^e=H_{\la ab\ra}$.
The fluid dynamics are given by the energy density $\rho$, 
the pressure
$p$, and the dissipative quantities $B$ (bulk viscous stress),
$Q_a$ (heat flux, $Q_au^a=0$) and $\pi_{ab}=\pi_{\la ab\ra}$
(shear viscous stress). These
arise in the energy-momentum tensor
\begin{equation}
T_{ab}=\rho u_au_b+(p+B)h_{ab}+2q_{(a}u_{b)}+\pi_{ab}\,,
\label{a}\end{equation}
where \cite{is}
\begin{equation}
q_a=Q_a+\left({\rho+p \over n}\right)j_a
\label{b}\end{equation}
is the total energy flux relative to $u^a$, with $n$ the particle
number density and $j_a$ the
particle flux ($j_au^a=0$). 
The latter are combined in the particle four-flow vector
\begin{equation}
N^a=nu^a+j^a \,.
\label{c}\end{equation}
In a self-consistent thermo-hydrodynamic description, we need
to introduce also the temperature $T$ and specific entropy $s$ per 
particle, defined in, or near to, equilibrium 
via the Gibbs equation
\begin{equation}
Tds=d\!\left({\rho\over n}\right)+p\,d\!\left({1\over n}\right)\,.
\label{d}\end{equation}

The hydrodynamic tensors $T_{ab}$ and $N^a$ define two natural
four-velocities -- the particle (or Eckart) four-velocity 
$u_{\rm p}^a$, for which $j_a=0\Leftrightarrow Q_a=q_a$,
and the energy (or Landau-Lifshitz) four-velocity
$u_{\rm e}^a$, for which $q_a=0\Leftrightarrow Q_a=-(\rho+p)j_a/n$.
These four-velocities coincide in equilibrium, and differ by a
small angle near to equilibrium.\footnote{It has recently been argued
\cite{kl} that only the energy frame is suitable for the description 
of irreversible thermodynamics.}
The four-velocity $u^a$ may be
chosen to be close to $u_{\rm p}^a$ and $u_{\rm e}^a$. Any
small change in $u^a$ produces second order changes (negligible
in the linear regime) in $\rho$, $p$, $n$, $T$, and $s$ \cite{is}.
These scalars therefore coincide (to first order) with the 
corresponding scalars for a local equilibrium state. The
bulk and shear viscous stresses $B$ and $\pi_{ab}$ are also
invariant to first order under a small change in $u^a$. Both
$q^a$ and $j^a$ undergo first-order changes, but the heat flux
vector $Q^a$ is invariant to first order. From Eq. (\ref{a}), we
see that
\begin{equation}
u^a=u_{\rm e}^a-{1\over(\rho+p)}q^a \,.
\label{d.}\end{equation}

For the isotropic and homogeneous Friedmann-Lemaitre-Robertson-Walker
(FLRW) universes
\begin{eqnarray*}
&&\D_a\theta=\D_a\rho=\D_ap=\D_aB=\D_an=\D_aT=\D_as=0\,, \\
&& \dot{u}_a=\omega_a=q_a=j_a=0\,, \\
&&\sigma_{ab}=E_{ab}=H_{ab}=\pi_{ab}=0\,.
\end{eqnarray*}
In covariant perturbation theory, a universe with small anisotropy
and inhomogeneity is characterized by these quantities being
small, and one neglects terms which are nonlinear in them. 
Since these quantities
vanish in the background, they are gauge-invariant \cite{eb}.
Note that FLRW models can admit scalar dissipation, in the form of
a bulk viscous stress $B$ (see, e.g., \cite{his,mm,z,zpm}),
reflecting the fact that
expanding fluids in general cannot maintain equilibrium \cite{is}.
However, we shall follow the standard approach in irreversible
thermodynamics of assuming an equilibrium background state, so that
$B=0$ in the background.\footnote{The consistency condition
that this assumption imposes through the transport equation (\ref{t1}) 
for $B$, is that the bulk viscosity $\zeta$ should be much less
than $\rho\theta^{-1}$.}
For convenience, the linearized Bianchi and Ricci equations that
underlie the covariant gauge-invariant theory are given in 
Appendix A (using the above notation and definitions, introduced 
in \cite{m2}, which considerably simplify the original equations).
Appendix A also contains useful differential identities.
Note that in the background 
\begin{equation}
\theta=3H\,,~\rho=3H^2(1+K)\,,~\dot{H}=-{\ts{1\over2}}H^2\left[ 
3(1+w)+(1+3w)K\right]\,,~B=0=\dot{s}\,,
\label{d'}\end{equation}
where $H=\dot{a}/a$ is the
Hubble rate, $a$ is the cosmic scale factor, $K=0,
\pm (aH)^{-2}$
is the dimensionless spatial curvature index, and $w=p/\rho$.

Linearization of the number conservation equation $\nabla_aN^a=0$ 
gives
\begin{equation}
\dot{n}+\theta n=-\D^aj_a \,.
\label{c'}\end{equation}
Using the energy conservation equation
(\ref{a7}) and the number conservation equation (\ref{c'}), 
together with Eq. (\ref{b}), the Gibbs equation (\ref{d}) implies
that 
\begin{equation}
nT\dot{s}=-3HB-\D^aQ_a \,.
\label{e}\end{equation}
The contribution of shear viscous stress to entropy generation is
via a nonlinear term $\sigma^{ab}\pi_{ab}$, so that
{\em in an almost-FLRW universe, the shear viscous stress does
not contribute to} $\dot{s}$. Thus non-dissipative
perturbations are not adequately characterized by $\dot{s}=0$. We
need to specify that $B=Q_a=\pi_{ab}=0$.

Scalar perturbations are covariantly and 
gauge-invariantly characterized by the spatial gradients of scalars. 
Density inhomogeneities are 
described by the comoving fractional density gradient \cite{eb}
\begin{equation}
\delta_a={a\D_a\rho\over\rho}\,.
\label{e'}\end{equation}
We define also the comoving expansion gradient \cite{eb},
and the dimensionless
fractional number density gradient (not considered in \cite{eb,bde}),
normalized pressure gradient, and normalized entropy gradient (see
\cite{not}) by
\begin{equation}
\theta_a=a\D_a\theta\,,~
\nu_a={a\D_an\over n}\,,~p_a={a\D_ap\over\rho}\,,~e_a=
{anT\D_as\over\rho}\,.
\label{e''}\end{equation}
Using the fact that $p=p(\rho,s)$, and 
the Gibbs equation (\ref{d}), we find
\begin{eqnarray}
p_a &=& \cs\delta_a+r e_a \,, \label{f'}\\
e_a &=& \delta_a-(1+w)\nu_a \,, \label{f''}
\end{eqnarray}
where the dimensionless quantities
\begin{equation}
\cs=\left({\p p\over\p\rho}\right)_{\!s},~~
r={1\over nT}\left({\p p\over\p s}\right)_{\!\rho},
\label{f.}\end{equation}
are respectively the adiabatic speed of sound and a non-barotropic 
index.
Note that in equations (\ref{f'}) and (\ref{f''}), these quantities 
and $w$ are 
evaluated in the background.\footnote{If $B$ is nonzero in 
the background, then
the background
speed of sound acquires an additional dissipative contribution
$c_{\rm b}$, where (see \cite{m})
$c_{\rm b}^{-2}=\beta_0(\rho+p)$, and $\beta_0$ arises 
in Eq. (\ref{t1}).}
In the background
\begin{equation}
\cs={\dot{p}\over\dot{\rho}}\,,~\dot{w}=-3H(1+w)(\cs-w)\,,
\label{f..}\end{equation}
where we have used $\dot{s}=0$ and 
the energy conservation equation (\ref{a7}).

A covariant thermodynamic classification of scalar perturbations 
is as 
follows. Perturbations are {\em non-dissipative} if 
$B=Q_a=\pi_{ab}=0$, and then in particular $\dot{s}=0$, so that the
specific entropy is constant along fluid flow-lines. If the specific
entropy is the same universal constant along all flow-lines, i.e.
if $e_a=0$ in addition to $\dot{s}=0$, then the perturbations
are {\em isentropic}, often (misleadingly) called `adiabatic'. 
For isentropic perturbations,
equations (\ref{f'}) and (\ref{f''}) show that
the number density perturbations and pressure perturbations are
algebraically determined by the energy density perturbations:
$\nu_a=\delta_a/(1+w)$, $p_a=\cs\delta_a$. 
The case of dissipative
perturbations with $e_a=0$, so that the specific entropy varies,
but only
along the fluid flow, will be called {\em dissipative
perturbations without entropy perturbations}. The
integrability condition $\dot{e}_a=0$, implies, via the
gradient of the entropy evolution equation (\ref{e}), that 
\begin{equation}
3H\D_aB+\D_a\left(\D^bQ_b\right)=0\,,
\label{h.}\end{equation}
which is very restrictive, except in the case where only shear viscous
stress is present.
In general, dissipative
perturbations will also involve entropy perturbations.

The evolution of the temperature is clearly affected by the nature
of the perturbations. In order to determine how this works, we
use $\rho$ and $s$ as the independent thermodynamic variables in the
Gibbs equation (\ref{d}). The integrability condition
$\p^2n/\p\rho\p s=\p^2n/\p s\p\rho$ then gives
\[
(\rho+p)\left({\p T\over\p\rho}\right)_{\!s}=\left(\cs+r\right)T \,,
\]
where we used
\[
\left({\p n\over\p s}\right)_{\!\rho}=-{n^2T\over \rho+p} \,,
\]
which follows from the Gibbs equation. 
Using the identity
\[
\dot{T}=\left({\p T\over\p\rho}\right)_{\!s}\dot{\rho}+
\left({\p T\over\p s}\right)_{\!\rho}\dot{s}\,,~
\]
together with the energy conservation equation (\ref{a7}) and
the entropy evolution equation (\ref{e}), the above equations lead to
the {\em temperature evolution equation}
\begin{eqnarray}
{\dot{T}\over T} &=& -\left(\cs+r\right)\theta
-{\left(\cs+r\right) \over \rho(1+w)} 
\left[3HB+\D^aq_a\right] \nonumber \\
{}&& -{1\over nT^2}\left({\p T\over\p s}\right)_{\!\rho}\left[3HB
+\D^aQ_a\right] \,. \label{h}
\end{eqnarray}
This equation reproduces the standard cooling rates for perfect fluids
in the nonrelativistic and ultra-relativistic cases.\footnote{A
similar equation is given in \cite{m}, but in the particle frame
only, and without separating out the non-barotropic $r$ terms.}
In the general case, the source terms on the right hand side 
show the role of non-barotropic and dissipative effects.
Note that the last term vanishes if the temperature is barotropic, 
i.e., if $T=T(\rho)$.
Bulk viscous perturbations counteract
the cooling due to expansion, shear viscous perturbations do not
affect the cooling rate (to first order), and the effect of
heat flux depends on the sign of the divergence.
For non-dissipative perturbations,
the sign of the non-barotropic index $r$ determines whether
cooling is enhanced or retarded.\footnote{Using 
the Gibbs equation (\ref{d}), we can show that $r=\cs[\rho(1+w)
\alpha-nc_{\rm p}]/nc_{\rm p}$, where 
$\alpha=n(\p n^{-1}/\p T)_p\geq0$ is the dilatation coefficient, and
$c_{\rm p}=T(\p s/\p T)_p\geq0$ is the specific heat at constant
pressure.}

We can also derive new evolution equations for the number density
perturbations, entropy perturbations, and temperature perturbations. 
The comoving gradient of
the number conservation equation (\ref{c'}), together with the
momentum conservation equation (\ref{a8}) and the
identity (\ref{a14}), gives
\begin{eqnarray}
\dot{\nu}_a+3r H\nu_a &=& -\theta_a+3(1+w)^{-1}\left(\cs+r\right)
H\delta_a \nonumber \\
{}&& {}+{a\over\rho(1+w)}\left[3H\left(\D_aB+\dot{q}_a+4Hq_a
+\D^b\pi_{ab}\right)+\D_a\D^b(Q_b-q_b)\right]\,.
\label{i}\end{eqnarray}
The comoving gradient of the entropy evolution equation (\ref{e}),
together with the energy conservation equation (\ref{a7}), the 
temperature evolution equation (\ref{h}), and the identity 
(\ref{a14}), gives the {\em evolution equation for 
entropy perturbations:}
\begin{equation}
\dot{e}+3H\left(\cs-w+r\right)e=-{a^2\over\rho}\left[3H\D^2B+
\D^2\left(\D^aQ_a\right)\right] \,,
\label{j}\end{equation}
where we have defined the scalar entropy perturbation
\begin{equation}
e=a\D^ae_a={a^2nT\over\rho} \D^2s \,.
\label{m}\end{equation}
Eq. (\ref{j}) shows that {\em for non-dissipative
or shear viscous perturbations, entropy perturbations
decay with expansion unless} $\cs-w+r\leq0$.
Defining the comoving fractional temperature gradient by
\begin{equation}
T_a={a\D_a T\over T}\,,
\label{h'}\end{equation}
we find from the evolution equation (\ref{h}) and the identity
(\ref{a14}), that the {\em evolution equation of 
(covariant and gauge-invariant) temperature perturbations} is
given by
\begin{eqnarray}
\dot{T}_a &=& -3\left(\cs+r\right)Ha\dot{u}_a-\left(\cs+r\right)
\theta_a-3Ha\D_a\left(\cs+r\right) \nonumber \\
&&{} -{a\left(\cs+r\right)\over\rho(1+w)}\left[3H\D_aB+\D_a
\left(\D^bq_b\right)\right]-{a\over nT^2}\left({\p T\over\p s}
\right)_{\!\rho}\left[3H\D_aB+\D_a\left(\D^bQ_b\right)\right]\,.
\label{h''}\end{eqnarray}

Now $\delta_a$ contains more information than just the
scalar density perturbations, since
at each point, $\delta_a$ picks out the direction of maximal 
inhomogeneity. The irreducible parts of the
comoving gradient of $\delta_a$ then describe completely 
and covariantly the variation in density inhomogeneities:
\begin{equation}
a\D_b\delta_a= {\ts{1\over3}}\delta\, h_{ab}+\xi_{ab}
+\ep_{abc}W^c\,,
\label{f}\end{equation}
where the scalar part $\delta\equiv a\D^a\delta_a=(a\D)^2\rho/\rho$ 
corresponds to the
usual gauge-invariant density perturbation scalar $\ep_{\rm m}$ 
\cite{b}, the vector part
$W_a=-{\ts{1\over2}}a\,\c\delta_a$ describes the rotational properties
of inhomogeneous clustering, and the tensor part
$\xi_{ab}=a\D_{\la a}\delta_{b\ra}$ describes the volume-true
distortion of inhomogeneous clustering. (These quantities 
were introduced in
\cite{bde}, but only the scalar $\delta$ was discussed.)
These irreducible parts encode respectively the total scalar, vector
and tensor contributions to density inhomogeneities.

It is difficult to
see how a rotation independent of the vorticity could arise,
and indeed we can show that $W_a$ is always proportional to the
vorticity vector:
\begin{equation}
W_a=-3a^2H(1+w)\omega_a \,.
\label{7}\end{equation}
This follows from the identity (\ref{a13}) and the energy conservation 
equation (\ref{a7}). Thus 
rotation in clustering matter is
inherited entirely (in the linear regime) from cosmic rotation:
{\em the vector
part of density inhomogeneities is determined completely in
direction by the
cosmic vorticity.}  The expansion and pressure index affect
the magnitude of the vector part. In particular, it follows
that $W_a=0$ if the background is non-expanding or De Sitter ($w=-1$).

The vorticity propagation equation (\ref{a2}) leads to
the new {\em evolution equation for the vector part of density 
inhomogeneities:}
\begin{equation}
\dot{W}_a+{\ts{1\over2}}H\left[3(1-w)+(1+3w)K\right]
W_a=-\left({3a^2H\over2\rho}\right)\c\left[\dot{q}_a+4Hq_a
+\D^b\pi_{ab}\right]\,,
\label{g}\end{equation}
where we have used equations (\ref{d'}) and
(\ref{f..}), and the momentum
conservation equation (\ref{a8}) allowed us to 
evaluate $\c\dot{u}_a$, together
with the identity (\ref{a13}). Unsurprisingly, Eq. (\ref{g})
shows that the scalar dissipative quantity $B$ does not influence
the evolution of the vector part of density 
inhomogeneities.\footnote{Although the gradient of $B$ occurs in
the transport equation (\ref{t2}) for $Q_a$, and therefore occurs
on the right hand side of (\ref{g}) in the particle frame ($q_a=Q_a$),
the curl of this
gradient is negligible by the identity (\ref{a13}), since $B$
vanishes in the background.} 
In the energy frame, heat flux also has no direct influence on $W_a$.
Eq. (\ref{g}) shows that {\em for non-dissipative
or only shear viscous perturbations, 
$W_a$ decays with expansion unless}
$3(1-w)+(1+3w)K \leq0$. For ordinary hydrodynamic matter, with
$0\leq w\leq {1\over3}$, this inequality is never satisfied
if the spatial curvature is non-negative.

To derive evolution equations for the tensor and scalar parts, we
take the comoving spatial gradient of the energy conservation equation
(\ref{a7}) and the Raychaudhuri equation (\ref{a1}), 
using the momentum conservation equation (\ref{a8}),
the identities (\ref{a19})--(\ref{a15}), and Eq. (\ref{7}):
\begin{eqnarray}
\dot{\delta}_a &=& 3wH\delta_a-(1+w)\theta_a 
 +{3aH\over\rho}\left[\dot{q}_a+4Hq_a+\D^b\pi_{ab}\right]
-{a\over\rho}\D_a\D^bq_b
\,, \label{14}\\
(1+w)\dot{\theta}_a &=& -2H(1+w)\theta_a-{\ts{3\over2}}H^2
\left[1+w+\left(1+w+{\ts{2\over3}}\cs\right)K\right]\delta_a 
-\cs\D^2\delta_a
\nonumber\\
{}&& -r\left(KH^2+\D^2\right)e_a
-{a\over\rho}\left(KH^2+\D^2\right)\D_aB 
\nonumber \\
{}&& +{3aH^2\over2\rho}\left[3(1+w)+(1+3w)K\right]
\left[\dot{q}_a+4Hq_a+\D^b\pi_{ab}\right] \nonumber \\
{}&& -{a\over\rho}\D_a\D^b\left[\dot{q}_b+4Hq_b+\D^c\pi_{bc}
\right]+{2\over a}\cs\, \c W_a
\,.\label{15}
\end{eqnarray}
Eq. (\ref{14}) can be shown to be 
in agreement with equation (61) of \cite{bde}, while Eq. (\ref{15}) 
generalizes equation (62) of \cite{bde} by including bulk viscous
effects.

We can now decouple the equations:
\begin{eqnarray}
&& \ddot{\delta}_a+H\left(2-6w+3\cs\right)\dot{\delta}_a
-{\ts{3\over2}}H^2\left[1+8w-3w^2-6\cs +
\left(1-3w^2+{\ts{2\over3}}\cs\right)K\right]\delta_a
\nonumber \\
&&{}~~~=\cs\D^2\delta_a-{2\over a}\cs \,\c W_a+r\left(KH^2
+\D^2\right)e_a+{a\over\rho}\left(KH^2+\D^2\right)\D_aB 
\nonumber \\
&&{}~~~+3{aH\over\rho}\left\{\ddot{q}_a+H\left[7-3w+3\cs
-(1+3w)K\right]\dot{q}_a \right.
\nonumber \\
&&{}~~~\left. +6H^2\left[1-3w+2\cs
-(1+3w)K\right]q_a-\cs \D_a\D^bq_b\right\}
\nonumber \\
&&{}~~~ +{a\over\rho}\left\{3H\D^b\dot{\pi}_{ab}+3H^2\left[2-3w+3
\cs-(1+3w)K\right]\D^b\pi_{ab}+\D_a\D^b\D^c\pi_{bc}
\right\} \,,
\label{k}\end{eqnarray}
where we have used equations (\ref{d'}), (\ref{e'}), (\ref{f'}),
and (\ref{a19})--(\ref{a15}).
The comoving gradient of the
evolution equation (\ref{k}) determines evolution 
equations for the scalar, vector and tensor parts of density
inhomogeneities, incorporating all dissipative and entropy
effects. We have already derived the vector evolution 
equation (\ref{g}). 
Taking the
comoving divergence of equation (\ref{k}), and using identities
(\ref{a19})--(\ref{a15}) and (\ref{a20}), gives
the {\em evolution equation for
scalar density perturbations}
\begin{eqnarray}
&& \ddot{\delta}+H\left(2-6w+3\cs\right)\dot{\delta}-{\ts{3\over2}}
H^2\left[1+8w-3w^2-6\cs+\left(1-3w^2+2\cs\right)K
\right]\delta-\cs\D^2\delta
\nonumber \\
&&{}~~~= {\sf S}[e]+{\sf S}[B]+{\sf S}[q]+{\sf S}[\pi]\,,
\label{l}\end{eqnarray}
where the source terms arising respectively from entropy
perturbations, bulk viscous stress, energy flux, and shear viscous
stress are:
\begin{eqnarray}
{\sf S}[e] &=&
r\left(3KH^2+\D^2\right)e \,, \label{l1}\\
{\sf S}[B] &=&\left(3KH^2+\D^2\right){\cal B} \,, \label{l2}\\
{\sf S}[q] &=&
3aH\left\{\ddot{q}+H\left[1-9w+\cs
-(1+3w)K\right]\dot{q}\right. \nonumber\\
&&\left.{}-{\ts{3\over2}}H^2\left[1+8w-9w^2-8\cs
+(1-9w^2)K\right]q-\cs \D^2q\right\} \,, \label{l3}\\
{\sf S}[\pi] &=&
3H\dot{{\cal S}}-3H^2\left[
1+6w-3\cs+(1+3w)K\right]{\cal S}+\D^2{\cal S}\,, \label{l4}
\end{eqnarray}
and we have defined the dimensionless perturbation scalars
\begin{equation}
{\cal B}={a^2\D^2B\over\rho}\,,~q={a\D^aq_a\over\rho}\,,~
{\cal S}={a^2\D^a\D^b\pi_{ab}\over\rho}\,.
\label{l5}\end{equation}
Eq. (\ref{l}) generalizes equation (74) of \cite{bde} to include bulk 
viscous effects, and is presented we believe in a more transparent
form, which makes clear the physical meaning of each term.

The new {\em evolution equation for the tensor part 
of density inhomogeneities}
follows from the trace-free symmetric part of the comoving gradient of
Eq. (\ref{k}), on using the identities (\ref{a19})--(\ref{a15}):
\begin{eqnarray}
&& \ddot{\xi}_{ab}+\left(2-6w+3\cs\right)H\dot{\xi}_{ab}
-{\ts{3\over2}}H^2
\left[1+8w-3w^2-6\cs+\left(1-3w^2+{\ts{2\over3}}\cs\right)K
\right]\xi_{ab}
\nonumber \\
&&{}- \cs\D_{\la a}\D_{b\ra}\delta
={\sf S}[e]_{ab}+{\sf S}[B]_{ab}+{\sf S}[q]_{ab}+
{\sf S}[\pi]_{ab}\,.
\label{n}\end{eqnarray}
The source terms are given by
\begin{eqnarray}
{\sf S}[e]_{ab} &=&
3rKaH^2\D_{\la a}e_{b\ra}+r\D_{\la a}\D_{b\ra}e \,,\label{n1}\\
{\sf S}[B]_{ab} &=&
3K{a^2H^2\over\rho}\D_{\la a}\D_{b\ra}B+\D_{\la a}
\D_{b\ra}{\cal B} \,, \label{n2}\\
{\sf S}[q]_{ab} &=&
3{a^2H\over\rho}\left\{\D_{\la a}\ddot{q}_{b\ra}+
H\left[7-3w+3\cs
-(1+3w)K\right]\D_{\la a}\dot{q}_{b\ra}  \right.
\nonumber \\
&&{} \left. 
+6H^2\left[1-3w+2\cs
-(1+3w)K\right]\D_{\la a}q_{b\ra}-\cs \D_{\la a}\D_{b\ra}
\D^cq_c\right\}\,, \label{n3} \\
{\sf S}[\pi]_{ab} &=&
{a^2\over\rho}\left\{3H\D_{\la a}\D^c\dot{\pi}_{b\ra c}
+3H^2\left[
2-3w+3\cs -(1+3w)K\right]\D_{\la a}\D^c\pi_{b\ra c} \right. \nonumber\\
&&{}\left.
+\D_{\la a}\D_{b\ra}\D^c\D^d\pi_{cd}\right\} \,.
\label{n4}\end{eqnarray}
Comparison of equations (\ref{l}) and (\ref{n}) shows that in the
simplest case of isentropic perturbations, the density distortion 
tensor $\xi_{ab}$ obeys the same equation as the scalar $\delta$,
so that 
\begin{equation}
\xi_{ab}= A_{ab}\,\delta\,,~~ \dot{A}_{ab}=0\,. 
\label{n5}\end{equation}
The
presence of entropy or dissipative perturbations breaks the simple
relation (\ref{n5}), 
and the evolution of the shape of density inhomogeneities 
is not directly determined by the scalar density perturbation.

\section{Causal transport equations}

We are now ready to introduce the evolution equations obeyed by
the dissipative quantities in the causal thermodynamics of
Israel and Stewart \cite{is}. 
This theory is based on a covariant treatment of the second law of
thermodynamics and the conservation equations, and its transport
equations are confirmed
by relativistic kinetic theory (via the relativistic 
generalization of the Grad approximation), which also provides
explicit expressions for the various thermodynamic parameters in
the case of a dilute gas. The theory thus has a firm physical
foundation. Furthermore, as pointed out earlier, dissipative signals
propagate below the speed of light and the equilibrium states
are stable, within the regime of validity of the theory. Thus the
causal and stable thermodynamics of Israel and Stewart is a
consistent relativistic thermodynamics which supercedes the
non-causal and unstable theories first put forward by Eckart
and Landau \& Lifshitz. 

The predictions of the causal theory agree
with those of the pathological theories in quasi-stationary
situations. But when high-frequency/ short-wavelength effects are
important, in the transient regime, the pathological theories are
inapplicable. Thus these theories cannot cover the full range of
behaviour of a relativistic fluid near equilibrium.
Moreover, these theories cannot even constitute 
part of a consistent
theoretical thermo-hydrodynamics because they are intrinsically
not relativistic theories, given their pathologies. Thus our
approach is to employ the causal thermodynamics to construct a 
self-consistent theory of cosmological density perturbations in the
general case. In particular applications, where it can be argued that 
the non-causal theories will give reasonable results, we can then
specialize the general equations appropriately. This is done in
the companion paper \cite{tm}.

The full form of the transport
equations, encompassing situations where the background equilibrium
state is accelerating and rotating, and including 
terms which were neglected in the original theory and restored
by Hiscock and Lindblom
\cite{hl}, is given in Appendix B for convenience.
Since we are dealing with cosmological perturbations, the background 
is non-rotating and non-accelerating, and spatial gradients of
thermodynamic coefficients give rise to nonlinear terms.
There are also linear terms, containing time derivatives of
thermodynamic coefficients, which were
restored by \cite{hl}. We will follow
the arguments of \cite{m,z} which show that under many reasonable
conditions, these terms may be neglected in
comparison with the other terms in the transport 
equations.\footnote{Note however that it is not always reasonable
to neglect these terms - see \cite{his,mm} for examples.}

With these simplifications, 
equations (\ref{b1})--(\ref{b3}) reduce to the {\em causal transport
equations}
\begin{eqnarray}
B &=& -\zeta \left[\theta+\beta_0\dot{B}-\alpha_0\D^aQ_a\right]\,,
\label{t1} \\
Q_a &=& -\kappa  \left[\D_aT+T\dot{u}_a+T\beta_1\dot{Q}_a
-T\alpha_0\D_aB-T\alpha_1\D^b\pi_{ab}\right] \,,
\label{t2}\\
\pi_{ab} &=& -2\eta\left[\sigma_{ab}+\beta_2\dot{\pi}_{ab}
-\alpha_1\D_{\la a}Q_{b\ra}\right]\,.
\label{t3}
\end{eqnarray}
The coefficients $\zeta$, $\kappa$, and $\eta$ of bulk viscosity,
thermal conductivity, and shear viscosity, appear also in the
non-causal (and the non-relativistic) theories. The coefficients
$\beta_I$ define characteristic relaxational time-scales 
\[
\tau_0=\zeta\beta_0\,,~\tau_1=\kappa T\beta_1\,,~\tau_2=2\eta
\beta_2\,,
\]
which are often taken to be of the order of the mean collision time,
but which are determined by collisional integrals in kinetic theory
\cite{is}. The non-causal theories are characterized by $\beta_I=0$.
Intuitively, this corresponds to instantaneous relaxation
to equilibrium when the
dissipative `force' is switched off. The coefficients $\alpha_I$,
which also vanish in the non-causal case, arise from a coupling
of viscous stress and heat flux (see Appendix B). They may also
be found from kinetic theory in the case of a dilute gas.
These transport equations hold in the energy and particle frames,
with suitable simple changes in some of the thermodynamic coefficients
(see Appendix B).

The transport equations (\ref{t1})--(\ref{t3}) are coupled to the
evolution equations for density inhomogeneities -- the scalar
equation (\ref{l}), the vector equation (\ref{g}), and the
tensor equation (\ref{n}). They are also coupled to the evolution
equations (\ref{i}), (\ref{j}) and (\ref{h''}) for 
number density, entropy, and temperature
perturbations. In all of these evolution equations, except the
entropy evolution equation (\ref{j}), 
the energy flux vector $q_a$ occurs. 
Considerable simplification is thus achieved by choosing the
energy frame ($q_a=0$), which is consistent with arguments in
favour of that frame \cite{kl,ho}.

In general, the coupling amongst the transport 
and evolution equations is highly complicated, 
although the coupled system can always be cast into a form
suitable for numerical integration.
Even in the non-dissipative case, when the transport equations
fall away, the evolution equations themselves are coupled.
In the simplest case of isentropic perturbations
($B=Q_a=\pi_{ab}=0$, $e=0$), the scalar
equation (\ref{l}) reduces to a wave equation only in $\delta$
(with undamped
phase speed $\cs$, as expected). In principle, the solution
of this equation, and the solution $W_a$ of Eq. (\ref{g}), may be used
to express Eq. (\ref{n}) as an equation only in $\xi_{ab}$. 

For
non-dissipative entropy perturbations, decoupling $\delta$ leads
to a third-order equation. First we apply the operator
$3KH^2+\D^2$ to Eq. (\ref{j}), using identity (\ref{a21}) and
\[
\dot K=H(1+3w)\left(1+K\right)K \,,
\]
to get
\[
\left[\left(3KH^2+\D^2\right)e\right]^{\rd}+H\left(2-3w+
3\cs+3r\right)\left[\left(3KH^2+\D^2\right)e\right]=0\,.
\]
Then we use Eq. (\ref{l}) in this latter equation, together
with identity (\ref{a21}) and Eq. (\ref{d'}), to obtain
the {\em decoupled density perturbation evolution equation for
non-dissipative entropy perturbations}
\begin{eqnarray}
&& \ddot{\delta}^{^{\rd}}+\left[\left(4-9w+6\cs+3r\right)H-
(\ln r)^{\rd}
\right]\ddot{\delta}-{\ts{1\over2}}AH\dot{\delta}-
{\ts{3\over2}}BH^2\delta \nonumber\\
&&{}~~~=\cs\D^2\dot{\delta}+\left[\left\{3\left(\cs-w+r\right)H-(\ln r)
^{\rd}\right\}\cs+\left(\cs\right)^{\rd}\right]\D^2\delta\,,
\label{d1}\end{eqnarray}
where $A$ and $B$ are complicated functions of $w,\cs,r,H,$ and 
$K$. In the case of a flat background $K=0$, we have
\begin{eqnarray*}
A &=&\left(1+84w-27w^2-69\cs+27w\cs-12r+36wr-18r\cs-
18c_{\rm s}^4\right)H
\nonumber\\
&&{}-6\left(\cs\right)^{\rd}+2\left(2-6w-3\cs\right)
(\ln r)^{\rd} \,,\\
B &=&\left(-1+10w-39w^2+24rw-15\cs+54w\cs+9w^2\cs
+3r-9w^2r-18r\cs-18c_{\rm s}^4\right)H \\
&&{}-6\left(\cs\right)^{\rd}-\left(1+8w-3w^2-6\cs\right)
(\ln r)^{\rd} \,.
\end{eqnarray*}

\subsection{The coupled system in general}

The complexity of Eq. (\ref{d1}) indicates the difficulty of
decoupling the equations for dissipative perturbations.
In general, the fact that the transport equations are
first-order in time derivatives shows that any decoupling will produce
at least one higher time derivative in the evolution equations.
In the non-causal limit ($\beta_I=0$), when the derivative terms
drop out of the transport equations, this does not hold, and the order
of the equations is the same as in the non-dissipative case.

The transport equations (\ref{t1})--(\ref{t3}) contain further
couplings amongst the dissipative quantities and couplings to
the density and entropy and temperature perturbations. These
further couplings are revealed when we take comoving spatial
gradients, and use the following expressions: 
\[
{a\D_a\zeta\over\rho}=\left({\p\zeta\over\p\rho}\right)_{\!s}\delta_a
+{1\over nT}\left({\p\zeta\over\p s}\right)_{\!\rho}e_a \,,
\]
which follows from equation (\ref{e''});
\[
a\D^a\theta_a={1\over 1+w}\left[3wH\delta-\dot{\delta}+3H{\cal S}
\right]\,,
\]
which follows from Eq. (\ref{14}), using the energy frame;
\[
a\D^a\dot{u}_a=-{1\over a(1+w)}\left[\cs\delta+re+{\cal B}+{\cal S}
\right] \,,
\]
which follows from the momentum conservation equation (\ref{a8});
\[
a^2\D^a\D^b\sigma_{ab}={\ts{2\over3}}a\D^a\theta_a \,,
\]
which follows from the constraint equation (\ref{a4}) and
the identity (\ref{a20});
and
\[
\D^a\D^b\D_{\la a}Q_{b\ra}=\left(\rho-3H^2\right)\D^aQ_a+
{\ts{2\over3}}\D^2\left(\D^aQ_a\right) \,,
\]
which follows from identity (\ref{a16}). Then operating on the
transport equations (\ref{t1})--(\ref{t3}) with, respectively,
$(a^2/\rho)\D^2$, $(a/\rho)\D^a$, and $(a^2/\rho)\D^a\D^b$, we get
the new {\em causal transport equations for the scalar dissipative
quantities}
\begin{eqnarray}
&& \tau_0\dot{{\cal B}}+\left[1-3(1+w)\tau_0H\right]{\cal B} =
 (\zeta\alpha_0 a)\D^2{\cal Q}+\left[{3\zeta H\over\rho(1+w)}\right]
 {\cal S} \nonumber\\
&&{}~~~
+{\zeta\over\rho(1+w)}\left[\dot{\delta}-3H\left\{w+
(1+w){\rho\over\zeta}\left({\p\zeta\over\p\rho}\right)_{\!s}\right\}
\delta\right] 
-\left[{3H\over nT}\left({\p\zeta\over\p s}
\right)_{\!\rho}\right]e \,,
\label{t4}\\
&& \tau_1\dot{{\cal Q}}+\left[1-3(1+w)\tau_1H\right]{\cal Q} =
-\left({\kappa T\over a\rho}\right){\cal T} \nonumber\\
&&{}~~~+{\kappa T\over a\rho(1+w)}\left[
\left\{\alpha_0\rho(1+w)-1\right\}{\cal B}
+\left\{\alpha_1\rho(1+w)-1\right\}{\cal S} \right]
+{\kappa T\over a\rho(1+w)}
\left[\cs\delta+re\right] \,, \label{t5}\\
&& \tau_2\dot{{\cal S}}+\left[1-H\left\{3(1+w)\tau_2-{4\eta\over 1+w}
\right\}\right]{\cal S} =\nonumber\\
&&{}~~~{\ts{2\over3}}\eta\alpha_1a\left(
\D^2{\cal Q}+3H^2K{\cal Q}\right) 
+{4\eta\over3(1+w)}\left[\dot{\delta}-
3wH\delta\right]
\,. \label{t6}
\end{eqnarray}
We have defined the scalars for heat flux and temperature
perturbations
\begin{equation}
{\cal Q}={a\D^aQ_a\over\rho} \,,~{\cal T}=a\D^aT_a \,.
\label{t7}\end{equation}
The 
comoving spatial divergence of equation (\ref{h''}) gives
the new {\em evolution equation for
scalar temperature perturbations:}
\begin{eqnarray}
\dot{{\cal T}} &=&\left({\cs+r\over 1+w}\right)\left[\dot{\delta}+ 3H
(\cs-w)\delta+3Hre\right]-3Ha^2\D^2(\cs+r) \nonumber\\
&&{}-{\rho\over nT^2}\left({\p T\over\p s}\right)_{\!\rho}
\left[3H{\cal B}+a\D^2{\cal Q}\right]
+3H\left({\cs+r\over 1+w}\right){\cal S}\,.
\label{t8}\end{eqnarray}

In summary, {\em the coupled system that governs scalar dissipative
perturbations in the general case} is given by:
the density perturbation equation (\ref{l}), the
entropy perturbation equation (\ref{j}), which we rewrite as
\begin{equation}
\dot{e}+3H\left(\cs-w+r\right)e=-3H{\cal B}-a\D^2{\cal Q}\,,
\label{t9}\end{equation}
and the equations (\ref{t4})--(\ref{t6}) and (\ref{t8}).

The number density perturbations do not occur in the coupled system.
Once $\delta$, ${\cal B}$, and ${\cal Q}$ are determined from the
coupled system, the scalar number density perturbations, defined
by $\nu=a\D^a\nu_a$, are found from the comoving divergence of
equation (\ref{i}). In the energy frame, this gives
\begin{equation}
\dot{\nu}+3rH\nu=(1+w)^{-1}\left[\dot{\delta}+3\left(\cs-w+r\right)H
\delta+3H{\cal B}+a^2\D^2{\cal Q}\right] \,.
\label{ndp}\end{equation}
The new evolution equation (\ref{ndp}) shows how bulk viscous
stress and heat flux govern the deviation of number density 
perturbations from energy density perturbations.
Note that shear viscous stress does not directly affect the number 
density perturbations.

For specific applications,
we present below the simplified coupled system that arises
in two special cases when only one form of dissipation is present. 

\subsection{Bulk viscous stress only}

The coupled system can be reduced to a pair of coupled equations in
$\delta$ (second-order in time) and $e$ (second-order in time).
In principle these may be decoupled. For a flat background, the
equations are:
\begin{eqnarray}
&& \ddot{\delta} 
+H\left(2-6w+3\cs \right)\dot{\delta} 
-{\ts{3\over2}}H^2\left(1+8w-3w^2-6\cs \right)\delta
\nonumber\\
&&{}~~~= \cs\D^2\delta+\left(w-\cs \right)\D^2e
-{1\over 3H}\D^2\dot{e} \,,
\label{d2}\end{eqnarray}
and
\begin{eqnarray}
&& \tau_0\ddot{e}+\left[1-{\ts{3\over2}}\left(1+3w-2\cs-2r\right)
\tau_0H\right]\dot{e}  \nonumber\\
&&{}-3H\left[w-\cs-r+3(1+w)r\tau_0H
-\tau_0\left(\cs+r\right)^{\rd}+{3H\over nT}\left({\p\zeta
\over\p s}\right)_{\!\rho}\right]e \nonumber\\
&&{}~~~=-\left[{\zeta\over H(1+w)}\right]
\dot{\delta}+{3\over(1+w)}\left[w\zeta+
\rho(1+w)\left({\p\zeta\over\p\rho}\right)_{\!s}\right]\delta \,.
\label{d3}\end{eqnarray}
Once these equations are solved for $\delta$ and $e$, the other
scalar quantities may be determined. Note that by the consistency
condition (\ref{h.}),
the entropy perturbations cannot be zero unless ${\cal B}$ itself
vanishes.

\subsection{Shear viscous stress only}

In the absence of entropy perturbations, the consistency
condition (\ref{h.})
is identically satisfied if only shear viscous stress is present, 
and the system reduces to the pair 
of coupled equations:
\begin{eqnarray}
&& \ddot{\delta}+H\left(2-6w+3\cs\right)\dot{\delta}-{\ts{3\over2}}
H^2\left[1+8w-3w^2-6\cs+\left(1-3w^2+2\cs\right)K
\right]\delta \nonumber\\
&&{}~~~=\cs\D^2\delta 
+3H\dot{{\cal S}}-3H^2\left[
1+6w-3\cs+(1+3w)K\right]{\cal S}+\D^2{\cal S}\,, \label{d5}\\
&&\tau_2\dot{{\cal S}}+\left[1-H\left\{3(1+w)\tau_2-{4\eta\over 1+w}
\right\}\right]{\cal S}\nonumber\\ 
&&{}~~~= 
{4\eta\over3(1+w)}\left[\dot{\delta}-
3wH\delta\right] \,. \label{d6}
\end{eqnarray}

Finally, we point out an interesting feature of the case when
only shear viscous stress is present (with or without
entropy perturbations).
The vector part $W_a$ of density inhomogeneities satisfies the
decoupled wave equation
\begin{eqnarray}
&&\tau_2\ddot{W}_a+\left[1+{\ts{3\over2}}\left\{1-w+(1+3w)K
\right\}\tau_2H\right]\dot{W}_a \nonumber\\
&&{}+{\ts{1\over4}}H\left[6(1-w)-(1+w)\left(3-w-6\cs\right)\tau_2H
+{24\eta\over\rho H} \right. \nonumber\\
&&\left.{}+\left\{2(1+3w)+\left[2(1+6w+3w^2)-
6(1+w)\cs+3(1+3w)^2K\right]\tau_2H \right.\right.\nonumber\\
&&\left.\left.{}
-{8\eta\over\rho H}
\left({1-3w\over 1+w}\right)\right\}K\right]W_a 
=\left[{\eta\over\rho(1+w)}\right]\D^2W_a\,.
\label{d4}\end{eqnarray}
The term $\D^b\sigma_{ab}$ that arises from the transport equation
(\ref{t3}) is eliminated via the constraint equation (\ref{a4}).
We used  the identity (\ref{a22}) and the constraint
equation (\ref{a5}) to evaluate $\c\c\omega_a$.
The undamped phase speed $v$ is clearly given by
\[
v^2={\eta\over\tau_2(\rho+p)} \,.
\]
It follows from the analysis of \cite{hl} that $v\leq 1$, as
expected from causality requirements.

\[ \]
{\bf Acknowledgements:}
JT thanks the Spanish Education Ministry for partial support 
under research project
number PB94-0718, and
the Department of Mathematics at Portsmouth for hospitality.

\newpage

\appendix
\section{Covariant propagation and constraint equations}

The Ricci identity for $u^a$ and the Bianchi identities 
(incorporating the field equations via the Ricci tensor), 
may be covariantly split into propagation and constraint equations 
(see \cite{bde}). In \cite{mes}, these equations are given in our
streamlined notation in the exact nonlinear case, for a perfect fluid.  
For an almost-FLRW universe, with imperfect fluid, the linearized 
form of the equations follows. (In the non-dissipative case, 
all right hand sides are zero.)\\

\begin{eqnarray}
\dot{\rho}+(\rho+p)\theta &=& -3HB -\D^aq_a \,,\label{a7}\\
(\rho+p)\dot{u}_a+\D_ap &=& 
-\D_aB-\dot{q}_a-4Hq_a-\D^b\pi_{ab} 
\,,\label{a8}\\
\dot{\theta}+{\ts{1\over3}}\theta^2-\D^a\dot{u}_a
+{\ts{1\over2}}(\rho+3p)
&=& -{\ts{3\over2}}B \,,
\label{a1}\\
\dot{\omega}_a+2H\omega_a
+{\ts{1\over2}}\c \dot{u}_a&=&0 \,,\label{a2}\\
\dot{\sigma}_{ab}+2H\sigma_{ab}
-\D_{\langle a}\dot{u}_{b\rangle }+E_{ab} &=& {\ts{1\over2}}\pi_{ab}
\,,\label{a3}\\
{\ts{2\over3}}\D_a\theta-\D^b\sigma_{ab}+\c\omega_a &=& q_a 
\,,\label{a4}\\
\D^a\omega_a &=&0 \,,\label{a5}\\
H_{ab}-\c\sigma_{ab}-\D_{\langle a}\omega_{b\rangle } &=& 0
\,,\label{a6} \\
\dot{E}_{ab}+3H E_{ab}-\c H_{ab}
+{\ts{1\over2}}(\rho+p)\sigma_{ab} &=&
-{\ts{1\over2}}\dot{\pi}_{ab}-{\ts{1\over2}}H\pi_{ab}-
{\ts{1\over2}}\D_{\la a}q_{b\ra}
  \,,\label{a9}\\
\dot{H}_{ab}+3H H_{ab}
+\c E_{ab} &=& {\ts{1\over2}}\c\pi_{ab}
 \,,\label{a10}\\
\D^bE_{ab}-{\ts{1\over3}}\D_a\rho &=& -Hq_a-{\ts{1\over2}}\D^b\pi_{ab}
 \,,\label{a11}\\
\D^bH_{ab}-(\rho+p)\omega_a &=& -{\ts{1\over2}}\c q_a
 \,.\label{a12}
\end{eqnarray}
\\

Some useful differential identities are \cite{eb,m2}
\begin{eqnarray}
\c\D_af &=&-2\dot{f}\omega_a \,,
\label{a13} \\
\D^2\left(\D_af\right) &=&\D_a\left(\D^2f\right) 
+{\ts{2\over3}}\left(\rho-3H^2\right)\D_a f+2\dot{f}\c\omega_a
\,, \label{a19}\\
\left(\D_af\right)^{\rd} &=& \D_a\dot{f}-H\D_af+\dot{f}\dot{u}_a \,,
\label{a14}\\
\left(a\D_aA_{b\cdots}\right)^{\rd} &=& a\D_a\dot{A}_{b\cdots}\,,
\label{a15}\\
\left(\D^2 f\right)^{\rd} &=& \D^2\dot{f}-2H\D^2 f 
+\dot{f}\D^a\dot{u}_a \,,\label{a21}\\
\D_{[a}\D_{b]}V_c &=& \left(H^2-{\ts{1\over3}}\rho\right)V_{[a}h_{b]c}
\,, \label{a16}\\
\D_{[a}\D_{b]}S^{cd} &=& 2\left(H^2-{\ts{1\over3}}\rho\right)
S_{[a}{}^{(c}h_{b]}{}^{d)} \,, \label{a17}\\
\D^a\c V_a &=& 0 \label{a20}\\
\D^b\c S_{ab} &=& {\ts{1\over2}}\c\left(\D^bS_{ab}\right)\,,
\label{a18}\\
\c\c V_a &=& \D_a\left(\D^bV_b\right)
-\D^2V_a+2\left({\ts{1\over3}}\rho-H^2\right)V_a \,,
\label{a22}\\
\c\c S_{ab} &=& {\ts{3\over2}}\D_{\la a}\D^cS_{b\ra c}-\D^2S_{ab}
+\left(\rho-3H^2\right)S_{ab} \,,
\label{a23}
\end{eqnarray}
where the vectors and tensors vanish in the background,
$S_{ab}=S_{\la ab\ra}$, and all the identities
except (\ref{a13}) are linearized.

\newpage

\section{Full causal transport equations}

For completeness and convenience, we amalgamate the results
of \cite{is} [equations (7.1)]\footnote{Note that the $a_1$ in 
equation (7.1c) of \cite{is} should be $\alpha_1$.}
and \cite{hl} [equations (21)--(23)]\footnote{Note that $\nabla_aq^a$ 
in equation (21) of \cite{hl} should be $q^{ab}\nabla_aq_b$ (using
their notation).}
in our notation, and present
the causal transport equations for viscous stress and heat flux in the
general (near equilibrium) case, covering both cosmological and
other (e.g., astrophysical) scenarios.
\begin{eqnarray}
B &=& -\zeta \left[\theta+\beta_0\dot{B}-\alpha_0\D^aQ_a\right]
\nonumber\\
&&{}+\zeta\left\{ a'_0\dot{u}^aQ_a+\gamma_0TQ^a\D_a\left({\alpha_0
\over T}\right)-{\ts{1\over2}}B\left[\beta_0\theta
+T\left({\beta_0\over T}\right)^{\rd}\right]\right\} \,,
\label{b1} \\
Q_a &=& -\kappa T \left[{1\over T}\D_aT+\dot{u}_a+\beta_1\dot{Q}_
{\la a\ra}-\alpha_0\D_aB-\alpha_1\D^b\pi_{ab}\right]
\nonumber \\
&&{}+\kappa T\left\{a_0B\dot{u}_a+a_1\dot{u}^b\pi_{ab}+\beta_1
\ep_{abc}Q^b\omega^c\right\}
\nonumber\\
&&{}+\kappa T\left\{(1-\gamma_0)TB\D_a\left({\alpha_0\over T}\right)
+(1-\gamma_1)T\pi_a{}^b\D_b\left({\alpha_1\over T}\right)-
{\ts{1\over2}}Q_a\left[\beta_1\theta+T\left({\beta_1\over T}\right)
^{\rd}\right]\right\} \,,
\label{b2}  \\
\pi_{ab} &=& -2\eta\left[\sigma_{ab}+\beta_2\dot{\pi}_{\la ab\ra}
-\alpha_1\D_{\la a}Q_{b\ra}\right]
\nonumber \\
&&{}+2\eta\left\{a'_1\dot{u}_{\la a}Q_{b\ra}+2\beta_2\ep_{cd(a}
\pi_{b)}{}^c\omega^d\right\}
\nonumber \\
&&{}+2\eta\left\{\gamma_1TQ_{\la a}\D_{b\ra}\left({\alpha_1\over T}
\right)-{\ts{1\over2}}\pi_{ab}\left[\beta_2\theta+T\left({\beta_2\over 
T}\right)^{\rd}\right]\right\} \,,
\label{b3}
\end{eqnarray}
where $\dot{Q}_{\la a\ra}\equiv h_a{}^b\dot{Q}_b$.
The coefficients $\zeta$, $\kappa$, and $\eta$ are, respectively,
the bulk viscosity, thermal conductivity, and shear viscosity. The
relaxation coefficients $\beta_0$, $\beta_1$, and $\beta_2$ 
are crucial to the causal behavior
of the theory. 
The coefficients $\alpha_0$ and $\alpha_1$ arise from a
coupling between viscous stress 
and heat flux, as reflected in the entropy 
four-current \cite{is}
\begin{eqnarray}
S^a &=& sN^a+{1\over T}Q^a+{\alpha_0\over T}B Q^a+
{\alpha_1\over T}\pi_{ab}Q^b
\nonumber \\
&&{}-{1\over2T}\left(\beta_0B^2+\beta_1Q_bQ^b+\beta_2\pi_{bc}
\pi^{bc}\right)u^a 
\nonumber\\
&&{}+{1\over 2T(\rho+p)}
\left( q^bq_b u^a+2\pi^{ab}q_b\right)\,.
\label{b4}
\end{eqnarray}
The coefficients $a_0$, $a'_0$, $a_1$, and $a'_1$ mediate the coupling
of acceleration and vorticity to viscous stress and heat flux, 
while $\gamma_0$ and
$\gamma_1$ appear due to a coupling of the spatial gradients of
$\alpha_I$ to viscous stress and heat flux. There are simple relations
between the unprimed and primed $a_I$ \cite{is}. 

A change from energy to particle frame results
in a re-definition of various thermodynamic coefficients, but
the transport equations maintain the same form.
A partial comparison is
given in \cite{is} (p. 350), but the heat flux equation in the
energy frame contains
the spatial gradient of the thermal potential
$\alpha=(\rho+p)/nT-s$, and not the temperature gradient or 
acceleration, which arise in the particle frame.
We can complete the comparison by
using the Gibbs equation (\ref{d}) and the momentum
conservation equation (\ref{a8}) to show that 
\begin{equation}
\left({nT\over\rho+p}\right)\D_a\alpha=-{1\over T}\D_aT-\dot{u}_a
+{1\over\rho+p}\left(\D_aB+\D^b\pi_{ab}\right)
\label{b5}\end{equation}
in the energy frame. It follows that the energy-frame equation 
(2.38b) of \cite{is} is in fact of the same form as the particle-frame
equation (2.41b). 

In the cosmological setting, all the terms representing coupling
of effects are nonlinear and may be neglected.
To first order, $\dot{Q}_{\la a\ra}=\dot{Q}_a$ and
$\dot{\pi}_{\la ab\ra}=\dot{\pi}_{ab}$.
Furthermore, we
follow the usual practice of neglecting the final term in square 
brackets on the right hand side of each transport equation. 
This can lead to problematic behavior in some cases, as shown in
\cite{his,mm} for the case of bulk viscosity. However, under a
range of reasonable conditions, these terms may be neglected in
comparison with the remaining terms \cite{z,m}. With these
assumptions, the terms in braces in the transport equations 
(\ref{b1})--(\ref{b3}) all
fall away, leading to the simplified cosmological transport equations
(\ref{t1})--(\ref{t3}).


\newpage

\begin{references}

\bibitem{b}
J. Bardeen, Phys. Rev. D {\bf 22}, 1882 (1980).

\bibitem{mfb}
V.F. Mukhanov, H.A. Feldman, and R.H. Brandenberger, Phys. Rep.
{\bf 215}, 203 (1992).

\bibitem{eb}
G.F.R. Ellis and M. Bruni, Phys. Rev. D {\bf 40}, 1804 (1989).

\bibitem{is} W. Israel and J. Stewart, Ann. Phys. (NY) {\bf 118},
341 (1979).

\bibitem{bde}
M. Bruni, P.K.S. Dunsby, and G.F.R. Ellis, Astrophys. J.
{\bf 395}, 54 (1992). 

\bibitem{hs}
W. Hu and N. Sugiyama, Astrophys. J. {\bf 444}, 489 (1995).

\bibitem{egs}
R. Maartens, G.F.R. Ellis, and W.R. Stoeger, Phys. Rev. D 
{\bf 51}, 1525 (1995).

\bibitem{tm}
J. Triginer and R. Maartens, in preparation.

\bibitem{hl}
W. Hiscock and L. Lindblom, Ann. Phys. (NY) {\bf 151}, 466 (1983).

\bibitem{d}
S.R. de Groot, W.A. van Leeuwen, and C.G. van Weert, 
{\em Relativistic Kinetic Theory} (North-Holland, Amsterdam, 1980).

\bibitem{caus}
M.A. Schweizer, Ann. Phys. (NY) {\bf 183}, 80 (1988);
D. Pavon, J. Bafaluy, and D. Jou, Class. Quantum Grav. {\bf 8},
347 (1991);
R. Narayan, A. Loeb, and P. Kumar, Astrophys. J. {\bf 431}, 
359 (1994);
J. Gariel and G. le Denmat, Phys. Lett. A {\bf 200}, 11 (1995);
A. Di Prisco, L. Herrera, and M. Esculpi, Class. Quantum Grav.
{\bf 13}, 1053 (1996);
J. Martinez, Phys. Rev. D {\bf 53}, 6921 (1996);
D. Jou, J. Casas-Vazquez, and G. Lebon, {\em Extended Irreversible
Thermodynamics} (Springer, Berlin, 1996).

\bibitem{his}
W. Hiscock and J. Salmonson, Phys. Rev. D {\bf 43}, 3249 (1991).

\bibitem{mm}
R. Maartens, Class. Quantum Grav. {\bf 12}, 1455 (1995),

\bibitem{z}
W. Zimdahl, Phys. Rev. D {\bf 53}, 5483 (1996).

\bibitem{w}
S. Weinberg, Astrophys. J. {\bf 168}, 175 (1971).

\bibitem{m2}
R. Maartens, Phys. Rev. D {\bf 55}, 463 (1997).

\bibitem{mes}
R. Maartens, G.F.R. Ellis, and S.T.C. Siklos, Class. Quantum 
Grav. {\bf 14}, 1927 (1997).

\bibitem{not}
We use units with $8\pi G=1=c$ and $k_{\rm B}=1$; $a,b,\cdots$ are
spacetime indices; (square) round brackets enclosing indices
denote (anti-) symmetrization, and angle brackets denote the
spatially projected, symmetric and trace-free part.
The main differences between our notation and that of \cite{bde}
are given by: $\D_a\equiv {}^{(3)}\nabla_a$, $\rho\equiv\mu$,
$\delta_a\equiv{\cal D}_a$, $\delta\equiv\Delta$,
$\theta_a\equiv{\cal Z}_a$, $p_a\equiv wP_a$, 
$e_a\equiv wnT{\cal E}_a/r$, $\xi_{ab}\equiv\Sigma_{ab}$.

\bibitem{zpm}
W. Zimdahl, D. Pavon, and R. Maartens, Phys. Rev. D. {\bf 55}, 
4681 (1997).

\bibitem{kl}
P. Kostadt and M. Liu, Preprint astro-ph/9610014 (1996).

\bibitem{m}
R. Maartens, in {\em Proceedings of Hanno Rund Workshop}, ed.
S.D. Maharaj (University of Natal, to appear) (astro-ph/9609119);
R. Maartens and J. Triginer, in preparation (1997).

\bibitem{ho}
W.A. Hiscock and T.S. Olson, Phys. Lett. A {\bf 141}, 125 (1989).


\end{references}
\end{document}